# OPTIMIZATION AND DYNAMIC CHARACTERIZATION SYSTEM PART IN TURNING

Claudiu Florinel BISU, Constantin ISPAS, Alain GERARD, Jean-Yves K'NEVEZ

*Abstract:* *The appearance of the self-sustained vibrations corresponds to the dynamic instability of the machine tool. An experimental device and a model were designed to characterize the cutting system, the block part. This analysis shows the need for knowing more details of it the behavior of the coin at the time of the cut, and for envisaging for the continuation the appearance of vibrations.*

*Key words: Block part, vibration, stiffness, natural frequency, displacement, turning*

## 1. INTRODUCTION

The appearance of vibrations during the operation of the machine tool cannot be avoided. Generally, these vibrations represent periodic displacements of the elastic system around its position of balance. The value of displacements depends as much on characteristic on the elements on the dynamic system as of the intensity on the interaction of these elements [1].

The elastic structure of the machine tool influences the stability of its dynamic system by the interaction with the process of cut. The causes of excitations of the system Part/Tool/Machine (**PTM**) can be directly related to machining or come from other sources, like defects of the kinematics chains, oscillations coming from the operation of the engines, the defects of balancing of the spindle, etc Nowadays, the machine tools are very rigid, and have geometrical defects less and less.

The dynamic problems (vibrations) are strongly related to the cut. These vibrations are generated and self-sustained by the process of cut. The self-sustained vibrations correspond to the dynamic instability of the machine tool [2].

## 2. DESCRIPTION OF EXPERIMENTAL PROTOCOL

To study these dynamic phenomena, we have a conventional turn having a high rigidity. At the time of an operation of cut (turning) with dynamic rate we represent system **PTM** by Fig.1. This system evokes the device of complete machining. This denomination gathers also the description of all the systems of connection. It takes into account the part that is interdependent of the machine, by the means of a device designed to ensure it's fixing. The elastic structure of the machine tool, system **PTM**, is a system with several degrees of freedom. It has a great number of naturals modes of vibration.

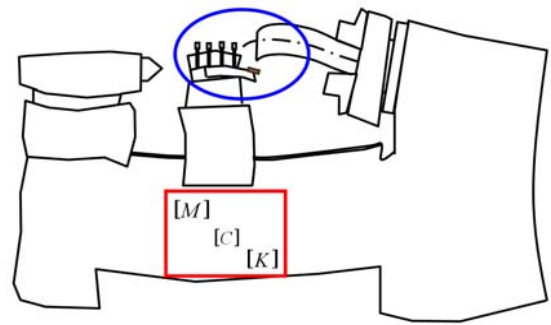

**Fig.1**. *Part-Tool-machine with dynamic rate of cut*

The vibratory behavior of each whole of the structure, is characterized by its natural frequency function of its stiffness, its mass and its damping coefficient. The elastic structure of system **PTM** is characterized by a high number of couples kinematics and sets with reduced tightening [3], [4]. To carry out the identification of the vibratory behavior of system **POM**, the machining system is divided into two blocks, the block tool (**BT**) and the block part (**BP**), Fig.2, [5].

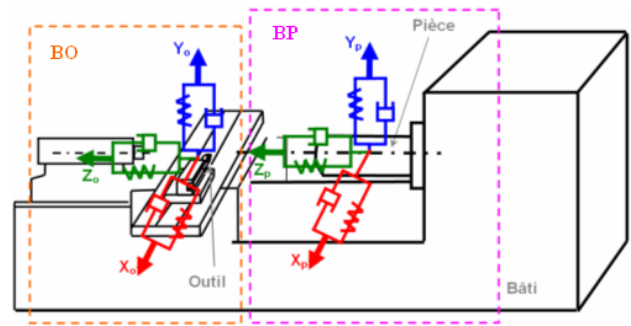

**Fig.2.** *Presentation of the experimental device*

## 3. BLOCK PART DEFINITION

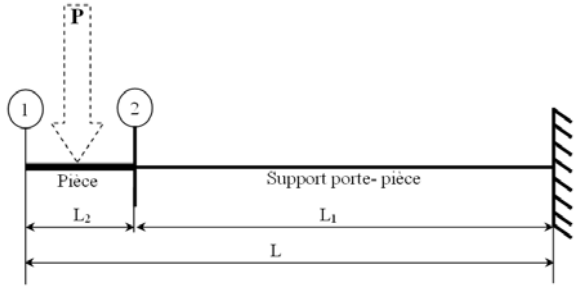

**Fig.3**. *Optimization of support part holder/part*

The part Block-Part (**BP**) is made up of the door coin, the coin and the pin. We consider the rigid coin, in front of the remainder of the elements of the **BT**. We thus conceive a unit (coin, part holder) very rigid, (3). The geometry being chosen, we must know the deformation of the coin at the time of an effort imposed, constant, according to the various lengths. With this intention, we dimension the coin by optimization of its stiffness. With the relations (1) and (2), we determine the values of the stiffness. These values are calculated for various lengths of support part holder/coin, with an effort P of *1000N*, a Young module $E= 2,1·10^5 N/mm^2$ and a moment of inertia $I$ determined for a diameter $D$ of 60 mm:

$$\delta = \frac{P \cdot L^3}{3EI} \quad (1)$$

$$I = \frac{\pi \cdot D^4}{64} \quad (2)$$

Represents the various stiffness according to the length of the support part holder. Taking into account the literature relating to the stiffness of the machine tools [1], we choose following dimensions:

*Table 1*
**Dimensioning of the coin is support coin.**

| D (mm) | L (mm) | $L_1$ (mm) | $L_2$ (mm) |
|---|---|---|---|
| 120 | 180 | 150 | 30 |

The geometry supplements selected to obtain a stiffness very important part is presented on.

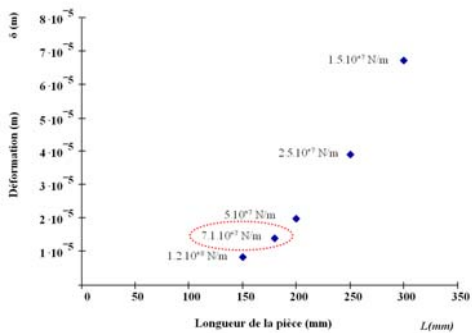

**Fig.4.** *Determination of the deformations for various lengths of support part holder*

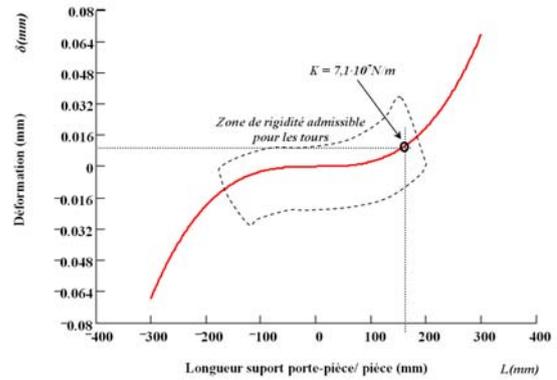

**Fig.5.** *Determination of the deformations for various lengths of support part holder.*

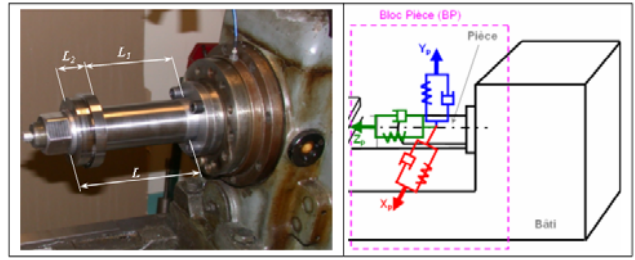

**Fig. 6.** *Representation of the BP part*

For these dimensions, the value of the stiffness is $7,1·10^7$ N/m. It represents the most important value of the beach of the stiffness of the turns, lain between $2-10·10^7$ N/m [6], [7] (Fig.4). With this design engineering, we carry out a configuration of part holder that confers a rigidity raised on the part. To integrate interaction **BT** and **BP** in modeling we must know the behavior of these two parts. To characterize the **BP**, we determine the masses, depreciation and the equivalent stiffness in the three directions

## 4. STATIC CHARACTERIZATION OF THE PART BLOCK PART (BP)

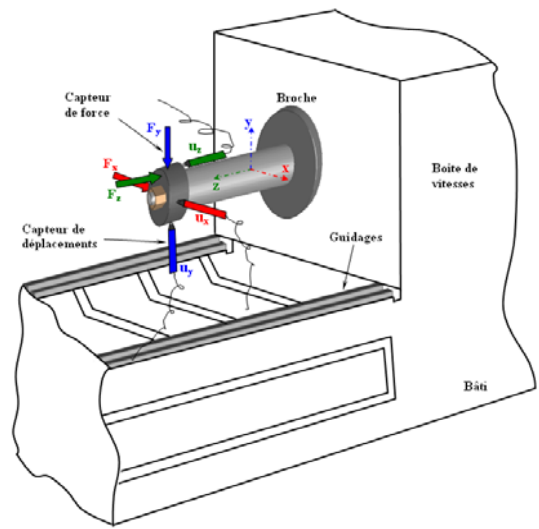

**Fig. 7.** *Experimental device for the static characterization of the BP*

The static tests consist in charging and measuring displacements of the **LP** following the three directions by determining the stiffness three-dimensional. The efforts applied are quantified using a sensor of force acting according to the three directions *x, y, z,* (respectively, radial axis, axis o f cut, center in advance). The torque of small displacements is measured by Fig. 3 sensors of one-way microphone-displacement. For the continuation we can determine the matrix of stiffness 3x3

The **BP** part has a very high rigidity and the couplings in this situation are very small compared to the value of the principal stiffness corresponding to the direction of load.

During the tests, the behavior of the **BP** proves to be linear, with an almost null hysteresis. Let us note that at the points of loading and unloading, the part is not influenced by the phenomenon of friction or the various plays generated by the assembled elements, as the pin or stages. We present the matrix of stiffness of the **BP** $[K_{BP}]$, the relation (3) during the tests of stiffness according to the three directions:

$$[K_{BP}] = \begin{bmatrix} 1,4 \cdot 10^7 & 0 & 0 \\ 0 & 2 \cdot 10^7 & 0 \\ 0 & 0 & 2.85 \cdot 10^8 \end{bmatrix} \quad (3)$$

We observe very clearly that the most important stiffness is located on Z, which is absolutely coherent with the configuration of the **BP**.

With an aim of knowing the dynamic behavior of the process of the cut, a dissociation between the behavior of the system and the process of cut are necessary. To carry out this decoupling, an experimental analysis detailed by impact was developed.

Our experimental device used for the procedure of tests by impact is detailed in., the **BP** part. Using a hammer with impact, the natural frequencies of each block, for each element, are given according to directions *x, y, z*. A three-axial accelerometer is positioned on each element subjected to the impact of the hammer. We carry out tests in each direction.

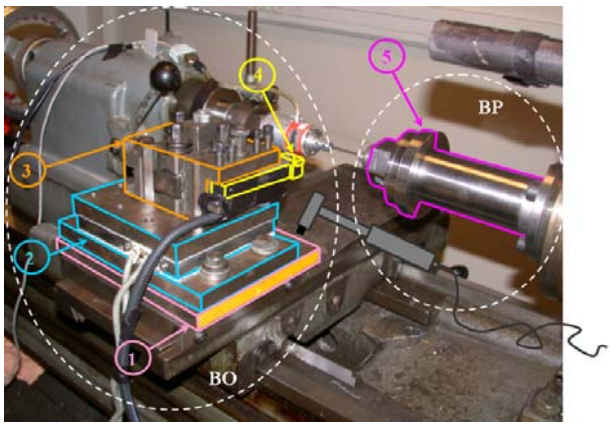

**Fig.8.** *Impact on BP*

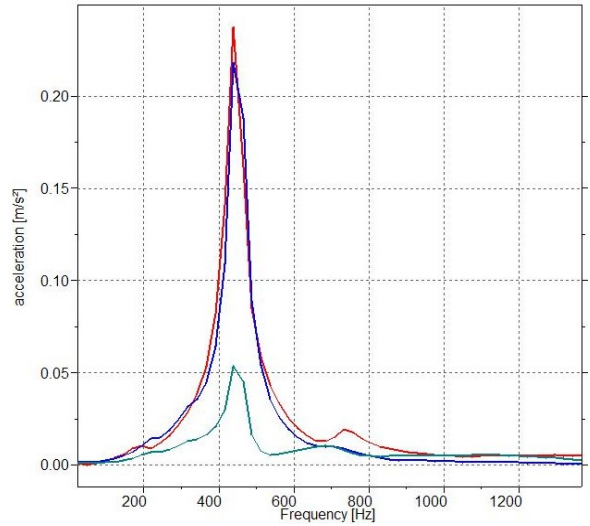

**Fig.9.** *Self frequencies for the BP part*

The results obtained are presented in 9. The beach of the naturals frequencies is represented for each element composing the system by carrying out a modal superposition.

The dynamic characterization of the machining system is supplemented by an analysis according to three configurations. To apply the modal superposition the three following configurations are analyzed: electric motor turning, electric motor turning with drive of the pin, electric motor turning with pin and movement coupled in advance.

The accelerometer 3D is placed on the body of the tool and the accelerometer 1D located on the frame side stitches. In 10, are presented the beaches of frequencies during the no-load, for the three quoted configurations. We notice that the frequencies in lower parts of 100Hz, which belong to the behavior of the engine, are frequencies of very low amplitude and which they appear on each configuration of tests. For example, the frequencies with tests for the three directions in the configuration "*electric motor turning with the drive of the pin and movement' advances coupled*" are traced in the figure below.

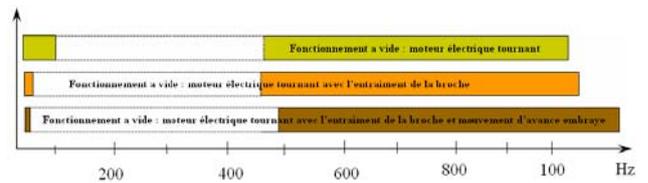

**Fig. 10**. *Representation of the beaches of frequency during the no-load of the machining system.*

An experimental device and a model were designed to characterize the machining system. Using the protocol of tests the matrix of stiffness 6x6 is obtained characterizing the system. Compared to other authors who use only the stiffness of the tool, this study takes into account, moreover, the stiffness resulting from the system of maintenance of the tool (**BT**), left which one does not occupy in this article [8].

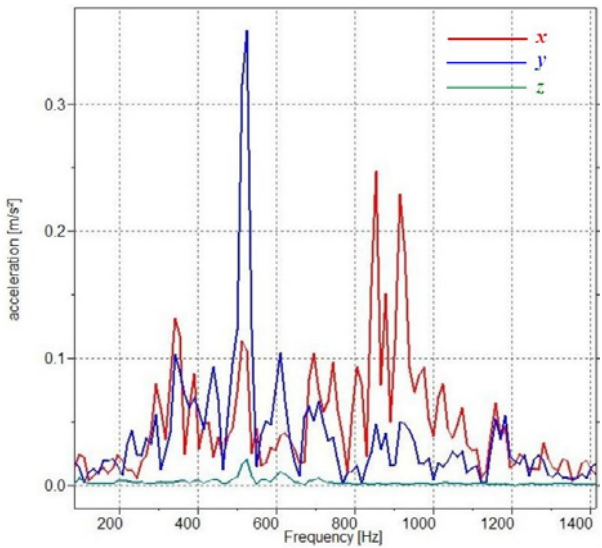

**Fig.11.** *Frequencies during the no-load in the third configuration, electric motor turning with drive of the pin and movement coupled in advance.*

Important information on the static stiffness is obtained, in particular making it possible to determine the privileged directions of displacement.
The center of stiffness and the center of rotation were in experiments given. The direction of minimum displacement was also defined starting from the experimental model. Then the **BP** part is characterized. Its stiffness are determined and a diagonal matrix of displacement is obtained.

The analysis of the static behavior of system **BP** will be validated thereafter by the analysis of the dynamic behavior at the time of the cut. The experimental protocol used in the determination of the static stiffness will be also validated by the tests of cut in the field of the vibratory cut.

During the tests, a dynamic analysis on the behavior of the system was carried out, in both cases of following tests: with the impact and during the no-load of the machine. Thus, we determined at the time of the impact the dynamic parameters, the matrix of mass, and damping stamps it.

## 5 CONCLUSIONS AND PERSPECTIVES

An experimental device and a model were designed to characterize the machining system. Using the protocol of tests the matrix of stiffness 3x3 is obtained characterizing the system part. The **BP** part is characterized. Its stiffness are determined and a diagonal matrix of displacement is obtained. During the tests, a dynamic analysis on the behavior of the system was carried out, in both cases of following tests: with the impact and during the no-load of the machine

The analysis of the static behavior of the system **BP** will be validated thereafter by the analysis of the dynamic behavior at the time of the cut.

The experimental protocol used in the determination of the static stiffness will be also validated by the tests of cut in the field of the vibratory cut.

This analysis shows the need for knowing more details of it the behavior of the part at the time of the cut, and for envisaging for the continuation the appearance of vibrations.

**Author(s):**

Claudiu-Florinel BISU, Doctorant, Université Bordeaux1, Laboratoire de Mécanique Physique UMR CNRS 5469, et LMSP, Université Politehnica de Bucarest, Splaiul Independentei 313 Bucarest – Roumanie, E-mail : cbisu@u-bordeaux1.fr
Constantin ISPAS, professeur des universités, Laboratoire de Machines et Systèmes de Production, Université Politehnica de Bucarest, Splaiul Independentei 313 Bucarest – Roumanie, E-mail : ispas1002000@yahoo.fr.
Alain GERARD, professeur des universités, Université Bordeaux 1, Laboratoire de Mécanique Physique UMR CNRS 5469, E-mail : alain.gerard@u-bordeaux1.fr
Jean-yves K'NEVEZ, maître de conférences, Université Bordeaux 1, Laboratoire de Mécanique Physique UMR CNRS 5469, E-mail : jean-yves.knevez@u-bordeaux1.fr